\documentclass{article} 
\usepackage{graphicx}
\graphicspath{%
    {converted_graphics/}% inserted by PCTeX
    {/}% inserted by PCTeX
}
\begin{document}

\begin{center}
{\LARGE An Improved Apparatus For Measuring}\vskip6pt

{\LARGE the Growth of Ice Crystals from Water Vapor}\vskip12pt

{\Large Kenneth G. Libbrecht}\vskip6pt

{\large Department of Physics, California Institute of Technology}\vskip3pt

{\large Pasadena, California 91125}\vskip1pt

\vskip18pt

\hrule\vskip1pt \hrule\vskip14pt
\end{center}

\textbf{Abstract. }We describe an apparatus designed for obtaining precise
measurements of the growth rates of ice crystals from water vapor over a
range of experimental conditions. Our aim is to produce clean, high-quality
test crystals in a well controlled environment for investigating the
detailed molecular dynamics that controls the basic physics of ice crystal
growth. In this paper we describe the nucleation and initial growth of test
crystals, their transport and selection into a experimental chamber, the
creation of a stable and controllable supersaturation, hardware and
calibration issues, and the crystal measurement via direct imaging and
broad-band interferometry.

\vskip8pt\noindent
[Note: The figures in this paper are presented at reduced resolution to facilitate rapid downloading. The paper is
available with higher quality figures at http://www.its.caltech.edu/˜atomic/publist /kglpub.htm, or by
contacting the author.]

\section{Investigating the Physics of Ice Crystal Growth}

\begin{figure}[htbp] % float placement: (h)ere, page (t)op, page (b)ottom, other (p)age
  \centering
  % file name: C:/1KGLaaa/aatempfold/VIGproject/Paper1-Apparatus/arxiv/Basic1.gif
  \includegraphics[bb=0 0 1004 448,width=4in,keepaspectratio]{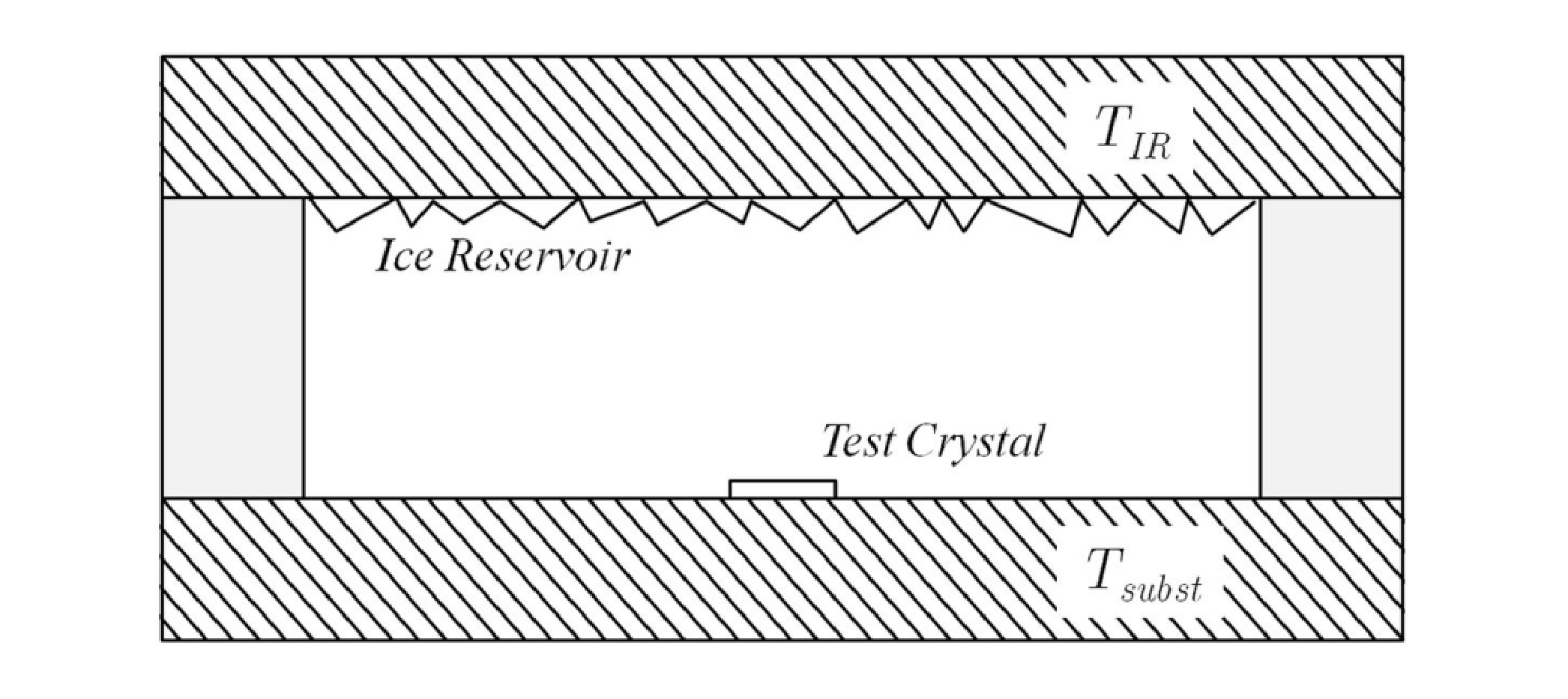}
  \caption{Idealized schematic of our
experimental set-up. An ice reservoir at temperature $T_{IR}$ supplies water
vapor for a test crystal resting on a substrate at temperature $T_{subst}.$
When $T_{IR}>T_{subst}$, growth rates are determined by measuring the size
and thickness of the test crystal as a function of time.}
  \label{basic}
\end{figure}

The goal of our ice growth experiments is to observe the growth of
individual ice crystals in a carefully controlled environment, and an
idealized schematic diagram of our experimental set-up is shown in Figure %
\ref{basic}. The top surface of the chamber is a thermal conductor with a
uniform temperature $T_{IR},$ and its inside surface is covered with a layer
of ice crystals that make up the ice reservoir. At the beginning of each
measurement a single test crystal is positioned near the center of the
bottom substrate surface held at temperature $T_{subst},$ separated from the
ice reservoir by thermally insulating side walls with a vertical spacing of
1.0 mm. The temperature difference $\Delta T=T_{IR}-T_{subst}$ determines
the effective supersaturation seen by the test crystal. After placing a test
crystal we then increase $\Delta T$ and observe its size and thickness as a
function of time, and from this extract growth velocities under various
conditions. Our ultimate goal is to understand the crystal growth dynamics
as a function of temperature, supersaturation, crystal morphology and
history, chemical make-up of the substrate, and the pressure and chemical
make-up of the background gas in the experimental chamber.

The means by which we produce these experimental conditions are described in
detail in this paper. As experience has shown repeatedly in investigations
of ice crystal growth, variations with regard to initial crystal nucleation,
growth history, and minute chemical impurities are inevitable in any
experimental system. As a result each individual crystal grows somewhat
differently, making it difficult to obtain experimental results that
describe theoretically perfect ice crystals. A good fraction of this paper
is therefore devoted to examining the absolute precision and reproducibility
in our measurements under a variety of conditions, and we have also strived
to produce quantitative measures of all processes that significantly affect
ice growth rates.

\subsection{Basic Growth Modeling}

It is instructive to first review the basic growth modeling of ice crystals
from water vapor, to define the notation used later. Following the formalism
in \cite{libbrechtreview} we write the growth velocity normal to the surface
as 
\begin{eqnarray}
v &=&\alpha \frac{c_{sat}}{c_{solid}}\sqrt{\frac{kT}{2\pi m}}\sigma _{surf}
\\
&=&\alpha v_{kin}\sigma _{surf}  \nonumber
\end{eqnarray}%
where the latter defines the kinetic velocity $v_{kin}.$ In this expression $%
kT$ is Boltzmann's constant times temperature, $m$ is the mass of a water
molecule, $c_{solid}=\rho _{ice}/m$ is the molecular number density for ice, 
$\sigma _{surf}=(c_{surf}-c_{sat})/c_{sat}$ is the supersaturation just
above the growing surface, $c_{surf}$ is the water vapor number density at
the surface, $c_{sat}(T)$ is the equilibrium number density above a flat ice
surface, and $\alpha \leq 1$ is the condensation coefficient.

For the simple case of a spherical crystal of radius $R,$ where we neglect
any temperature variations from latent heat generation, a straightforward
solution of the particle diffusion equation \cite{libbrechtreview} yields
the growth velocity 
\begin{eqnarray}
v &=&\frac{\alpha \alpha _{diff}}{\alpha +\alpha _{diff}}v_{kin}\sigma
_{\infty }  \label{diff2} \\
&=&\frac{\alpha }{\alpha +\alpha _{diff}}\frac{c_{sat}D\sigma _{\infty }}{%
c_{solid}R}  \nonumber
\end{eqnarray}%
where $\sigma _{\infty }$ is the supersaturation far from the growing
crystal, $D$ is the diffusion constant, $R$ is the sphere radius, and%
\begin{eqnarray}
\alpha _{diff} &=&\frac{c_{sat}D}{c_{solid}v_{kin}R}=\frac{D}{R}\sqrt{\frac{%
2\pi m}{kT}}  \label{diff3} \\
&\approx &0.15\left( \frac{1\ \mu \textrm{m}}{R}\right) \left( \frac{D}{%
D_{air}}\right)  \nonumber
\end{eqnarray}%
where the latter expression is evaluated for the specific case of ice
growing at $T=-15$ C in air. At a pressure of one atmosphere, $%
D=D_{air}\approx 2\times 10^{-5}$ m$^{2}/\sec ,$ while at lower pressures $%
D\sim P^{-1}.$

We note two limiting cases in Equation \ref{diff2}. When $\alpha \ll \alpha
_{diff},$ then $\sigma _{surf}\approx \sigma _{\infty }$ and $v\approx
\alpha v_{kin}\sigma _{\infty }$, which describes purely kinetics-limited
growth unhindered by particle transport. In the opposite limit, $\alpha
_{diff}\ll \alpha ,$ we have $\alpha (\sigma _{surf})\sigma _{surf}\approx
\alpha _{diff}\sigma _{\infty }$ and $v\approx \alpha _{diff}v_{kin}\sigma
_{\infty },$ which describes purely diffusion-limited growth. As we will see
below, we have found that that Equation \ref{diff2} with $\alpha _{diff}$ as
a simple adjustable constant is a useful approximation in many of our
experiments.

\subsubsection{Crystal Heating Effects}

An analysis along the same lines as the above can also be used to show the
approximate effects of crystal heating generated by solidification in our
experiments. At low background pressures, the growth velocity of a
hemispherical crystal in thermal contact with the substrate can be written 
\cite{libbrechtreview}

\begin{equation}
v=\frac{\alpha \alpha _{cond}}{\alpha +\alpha _{cond}}v_{kin}\sigma _{\infty
}  \label{heating}
\end{equation}%
with 
\begin{equation}
\alpha _{cond}\approx 25G\left( \frac{1\ \mu \textrm{m}}{R}\right)
\end{equation}%
where $G\approx 1$. We note that $\alpha _{cond}\approx \alpha _{diff}$ at a
background pressure of $P\approx 5$ Torr; for higher pressures the effects
of crystal heating can be ignored and the growth is limited mainly by
particle diffusion and attachment kinetics. Since Equations \ref{diff2} and %
\ref{heating} have similar functional forms, the overall effects of particle
diffusion and heating on the measured crystal growth velocities $v(\sigma
_{\infty })$ are also expected to show similar functional forms.

We also note that the temperature difference between the growing top surface
of an ice crystal plate and the lower surface (in contact with the
substrate) is approximately%
\begin{eqnarray*}
\Delta T &\approx &\frac{hv\rho \lambda }{\kappa } \\
&\approx &0.05\left( \frac{h}{50\ \mu \textrm{m}}\right) \left( \frac{v}{%
1\ \mu \textrm{m/sec}}\right) \textrm{ C}
\end{eqnarray*}%
where $h$ is the plate thickness, $v$ is the growth velocity, $\rho =917$
kg/m$^{3}$ is the density of ice, $\lambda =2.8\times 10^{6}$ J/kg is the
latent heat of solidification from water vapor, and $\kappa =2.3$ Wm$^{-1}$K$%
^{-1}$ is the thermal conductivity of ice. In an absolute sense, this
difference is also expected to be negligible in our experiments.

\section{A Series of Ice Growth Chambers}

A significant challenge in obtaining accurate ice crystal growth
measurements is the production of high-quality test crystals in a well
controlled environment. Ideally these crystals should be clean (free from
chemical impurities on the ice surface), small (no larger than 50 $\mu $m),
free from significant dislocations, isolated at a convenient location on a
substrate, and with a well-defined crystal orientation relative to the
substrate surface. To accomplish this we have chosen to create free-falling
crystals in a large outer chamber and then transport and select crystals in
a second, inner chamber. Within the inner chamber, crystals are further
placed within a smaller sub-chamber, in which the supersaturation and
temperature are carefully controlled. This strategy separates various
experimental functions - nucleating crystals, selecting crystals, and
measuring growth rates - into different physical regions, so each function
can be better optimized and controlled. We now describe this series of
growth chambers in more detail. A number of features of this new apparatus
have been adapted and improved from previous versions described in \cite%
{ver1} and \cite{ver2}. A review of earlier related experimental work is
given in \cite{libbrechtreview,critical,kkreview}.

\subsection{The Outer Chamber}

\begin{figure}[htbp] % float placement: (h)ere, page (t)op, page (b)ottom, other (p)age
  \centering
  % file name: C:/1KGLaaa/aatempfold/VIGproject/Paper1-Apparatus/arxiv/outerchamber.gif
  \includegraphics[bb=0 0 1338 1215,width=4.3in,keepaspectratio]{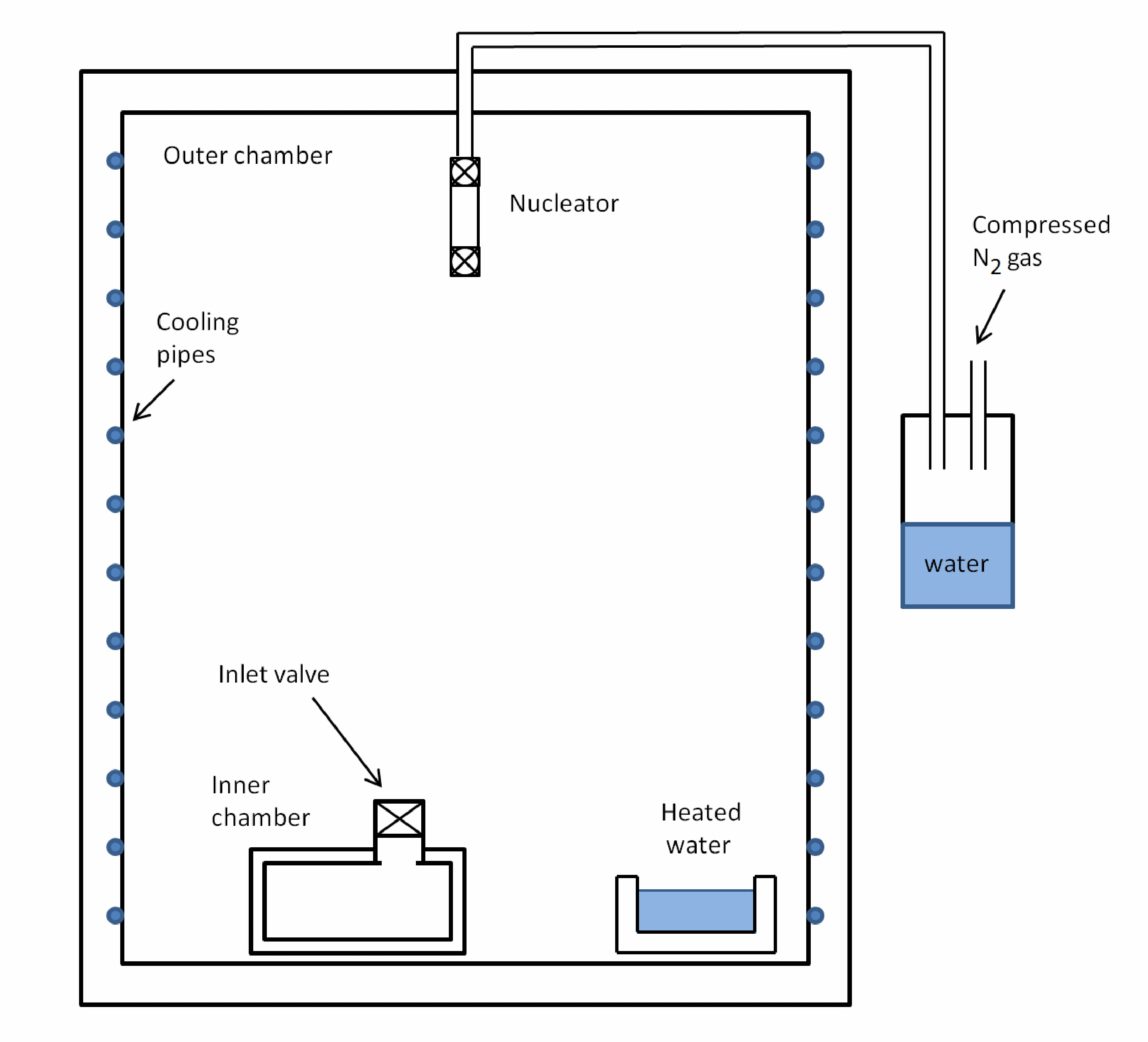}
  \caption{A schematic view (not all parts
to scale) of the outer and inner ice crystal growth chambers, as described
in the text. Ice crystals are nucleated every ten seconds near the top of
the outer chamber, after which they grow in the supersaturated air for
several minutes until they become heavy enough to fall to the bottom of the
chamber. When desired, an inlet valve is opened to admit crystals into the
inner chamber, where their growth rates are measured under controlled
conditions.}
  \label{chamber1}
\end{figure}

Ice crystals are initially created in air within a convection chamber shown
schematically in Figure \ref{chamber1}, measuring 90 cm in height and 50x50
cm in width and depth (inside dimensions). A programmable chiller cools the
copper walls of the chamber by circulating methanol through soldered cooling
pipes. Once the system is stable, the chiller can maintain the interior
temperature of the chamber down to -35 C with an accuracy of about 0.1 C.

An insulated reservoir containing heated water located on the bottom of the
outer chamber introduces water vapor into the chamber via continuous
evaporation. The chamber contains ordinary laboratory air at a pressure of
one atmosphere, so convection transports and mixes the water vapor into the
air, resulting in a steady-state water vapor supersaturation within the
chamber. We have found that the temperature and supersaturation are
surprisingly uniform within the chamber \cite{morrison}, indicating
efficient mixing by convection. The spatial profile and temporal variability
of supersaturation within the chamber are difficult to determine accurately,
although these both likely increase monotonically with water temperature 
\cite{morrison}. Since the function of the outer chamber is mainly to
produce candidate ice crystals, it is not necessary to know the
environmental conditions in this region with high accuracy. We typically
heat the water to approximately 25 C, and we typically operate with a
central temperature of approximately -12.5 C when the goal is to produce
thin hexagonal plate crystals.

A pulse of rapidly expanding gas is used to nucleate the growth of ice
crystals near the top of the outer chamber \cite{hallett,morrison}. The
nucleator is made from a 5-cm-long pipe, 2 cm in diameter, with
solenoid-actuated valves on both ends, connected to a source of compressed
gas that has been saturated with water vapor from a room-temperature water
reservoir, as shown in Figure \ref{chamber1}. The first valve is opened for
about five seconds to admit compressed gas into the pipe, and then this
valve is closed. The second valve is then opened to discharge the compressed
gas into the growth chamber. The rapid expansion cools the saturated gas
inside the pipe to nucleate small ice crystals \cite{hallett,morrison}. The
presence of these faceted crystals is easily verified by observing the
sparkle from a bright flashlight shining into the chamber. We typically use
nitrogen gas at 20 psi in the nucleator, although argon, helium, and other
gases work as well. The nucleator valves are cycled every ten seconds, thus
yielding a steady-state of small crystals growing and falling inside the
outer chamber.

We estimate that under typical conditions roughly $10^{7}-10^{8}$ ice
crystals are present in the chamber at any given time, and each grows for
about three minutes before gravity causes it to fall to the chamber bottom.
The typical size of an ice crystal plate when it reaches the chamber floor
is a few microns thick and \symbol{126}20 microns in diameter. Because there
is a continuous downward flow of ice crystals in the chamber, the system as
a whole is somewhat self-cleaning. Chemical impurities and dust particles
within the chamber are incorporated into growing crystals, which fall and
are thus removed. The overall result is that we have a continuous source of
rather pristine, newly formed, ice crystals. When desired, an inlet valve
(see Figure \ref{chamber1}) is opened to admit a random sample of these
crystals into the inner chamber for further experimentation.

When the outer chamber is at -12.5 C, we have found that we can transfer
crystals to the inner chamber only when its temperature is below -4 C. At
higher temperatures, crystals apparently evaporate before reaching the
substrate. To take data at higher temperatures we therefore transfer at -4 C
and subsequently raise the temperature of the inner chamber to the desired
operating temperature. The latter step must be done slowly at a pressure of
one atmosphere, while making sure the ice reservoir is kept in equilibrium
with the test crystal while the temperature changes.

\subsection{The Inner Chamber}

\begin{figure}[htbp] % float placement: (h)ere, page (t)op, page (b)ottom, other (p)age
  \centering
  % file name: C:/1KGLaaa/aatempfold/VIGproject/Paper1-Apparatus/arxiv/innerchamber.gif
  \includegraphics[bb=0 0 1567 927,width=4.6in,keepaspectratio]{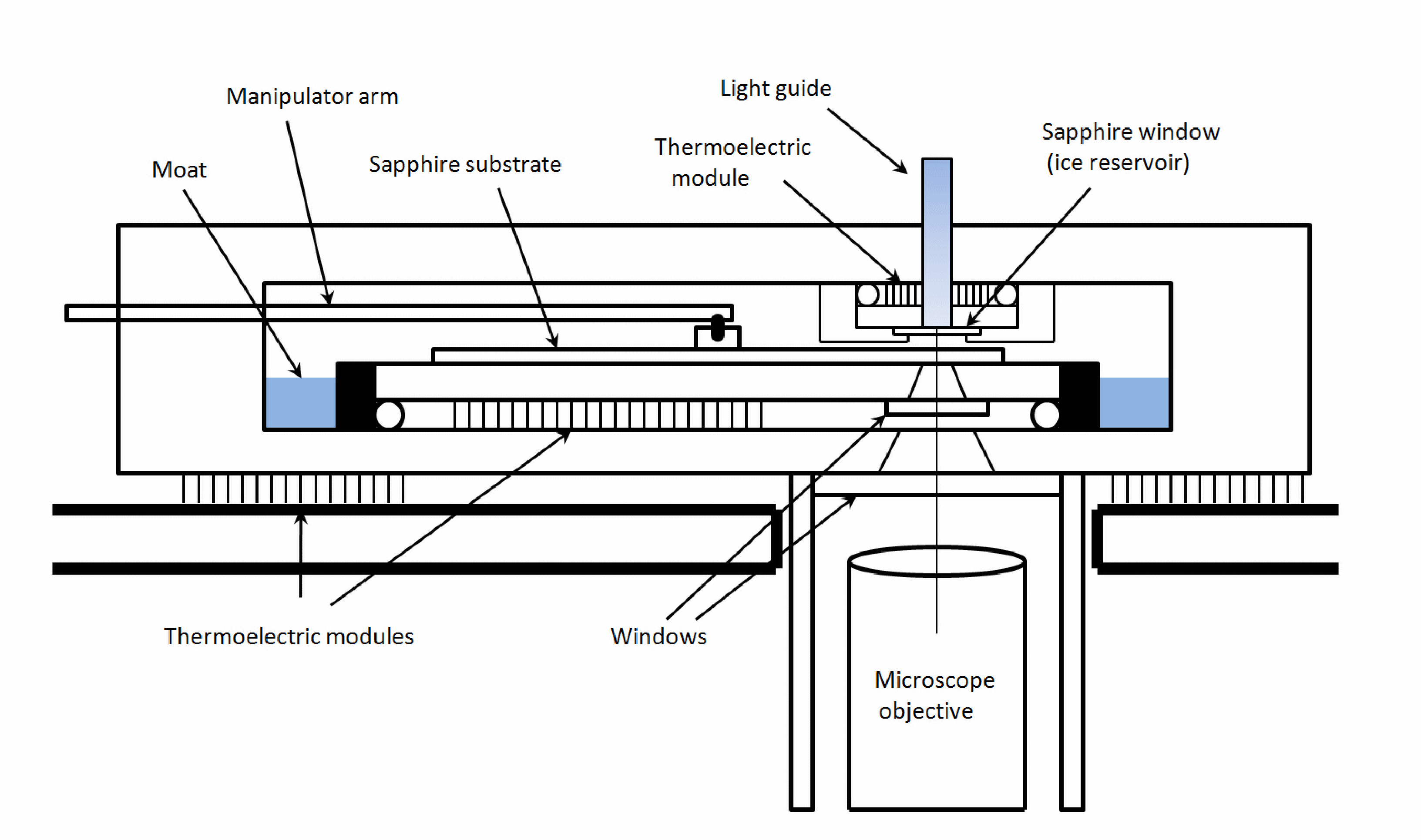}
  \caption{A schematic view (not all parts
to scale) of the inner growth chamber, as described in the text. After being
admitted through an inlet valve (not shown here; see Figure \protect\ref%
{chamber1}), ice crystals are selected and positioned within an observing
region (the experimental sub-chamber described in the text) for subsequent
growth measurements.}
  \label{innerchamber}
\end{figure}

Figure \ref{innerchamber} shows a schematic view of the inner growth
chamber, within which ice crystal growth measurements are performed. Air
from the outer chamber is drawn into the inner chamber via the inlet valve
(shown in Figure \ref{chamber1}), depositing a sample of ice crystals onto
the sapphire substrate -- a disk 50.8 mm in diameter and 1.0 mm thick, with
its optic axis oriented along the disk axis (to reduce birefringent
effects). Plate-like crystals typically land on the substrate with one basal
face in contact with the surface, thus orienting the ice crystal axis
relative to the substrate surface. By rotating this disk about its central
axis (using a small silicone o-ring in contact with the outer edge of the
disk), and translating it via manipulator arm (shown in Figure \ref%
{innerchamber}), it is straightforward to find a suitable ice crystal and
position it near the optical axis of the microscope objective for performing
subsequent growth measurements. When taking data our goal is usually to
position one and only one crystal at the center of the test region, to
ensure a well-defined supersaturation during the measurement process.

The inner chamber is a small vacuum chamber constructed to keep the interior
as free from vaporous chemical contaminants as is feasible. The chamber body
is made from anodized aluminum, as are several interior parts. The
temperature-control elements, in particular the greased thermoelectric
modules shown in Figure \ref{innerchamber} and their associated thermistors
and wiring, are all mounted outside the vacuum envelope, separated by
silicone o-rings. Other materials inside the chamber include sapphire,
coated optical windows, polycarbonate, stainless steel, and a small amount
of vacuum-compatible epoxy and vacuum grease (Apiezon N). Several hoses made
from polyflow tubing and short lengths of silicone tubing are connected to
the inner chamber via stainless steel tubing. The inlet valve is a 1.33-inch
stainless steel vacuum butterfly valve retrofit with silicone o-rings. The
sapphire substrate is typically cleaned with isopropyl alcohol and rinsed
with deionized water between runs, and the water in the moat is removed and
replaced between runs as well. The inner chamber is also partially cleaned
and baked at 30 C between runs, and the air inside is frequently replaced
during a run. With these precautions we believe that chemical influences on
our growth measurements are fairly low \cite{chemical1}. Nevertheless,
chemical influences cannot be excluded completely, so this remains a
potentially significant systematic error in our growth measurements.

The inner chamber base, the chamber lid, and the substrate base are
separately temperature regulated, using thermistor sensors that have an
absolute accuracy of better than 0.1 C. Typically the three temperature
set-points are identical, producing a nearly isothermal chamber. The
temperature of the upper sapphire window (the ice reservoir shown in Figure %
\ref{innerchamber}) is then separately temperature regulated to control the
supersaturation seen by the test crystal under observation.

Special care was taken in the temperature regulation of the ice reservoir,
in order to achieve both high stability and tunability of the
supersaturation. We adapted the temperature controller described in \cite%
{tempcontroller} for this purpose, adding an additional layer of tunability
to achieve very fine set point control. We also read out the temperature
controller output voltage $V_{IR}(T_{IR})$ using a Keithley precision
voltmeter that gave microvolt stability over extended periods of time.
Calibration and overall accuracy of this temperature control is discussed
below.

The inner chamber includes external tubing connections to a vacuum pump, a
vacuum gauge, and a gas inlet. The ice moat shown in Figure \ref%
{innerchamber} serves the purpose of keeping the interior of the chamber
saturated with water vapor after repeated pump-downs, and at a range of
pressures. A diaphragm pump can produce pressures as low as 1 Torr inside
the chamber. The microscope objective is at a temperature near room
temperature, with two windows and two aluminum thermal baffles separating it
from the substrate. A flow of 2 cc/sec of dry nitrogen gas into the
partially sealed microscope objective cell is sufficient to prevent
condensation on the cold window surface under typical operating conditions.

\subsection{The Experimental Sub-Chamber}

\begin{figure}[htbp] % float placement: (h)ere, page (t)op, page (b)ottom, other (p)age
  \centering
  % file name: C:/1KGLaaa/aatempfold/VIGproject/Paper1-Apparatus/arxiv/subchamber.gif
  \includegraphics[bb=0 0 1281 570,width=4.5in,keepaspectratio]{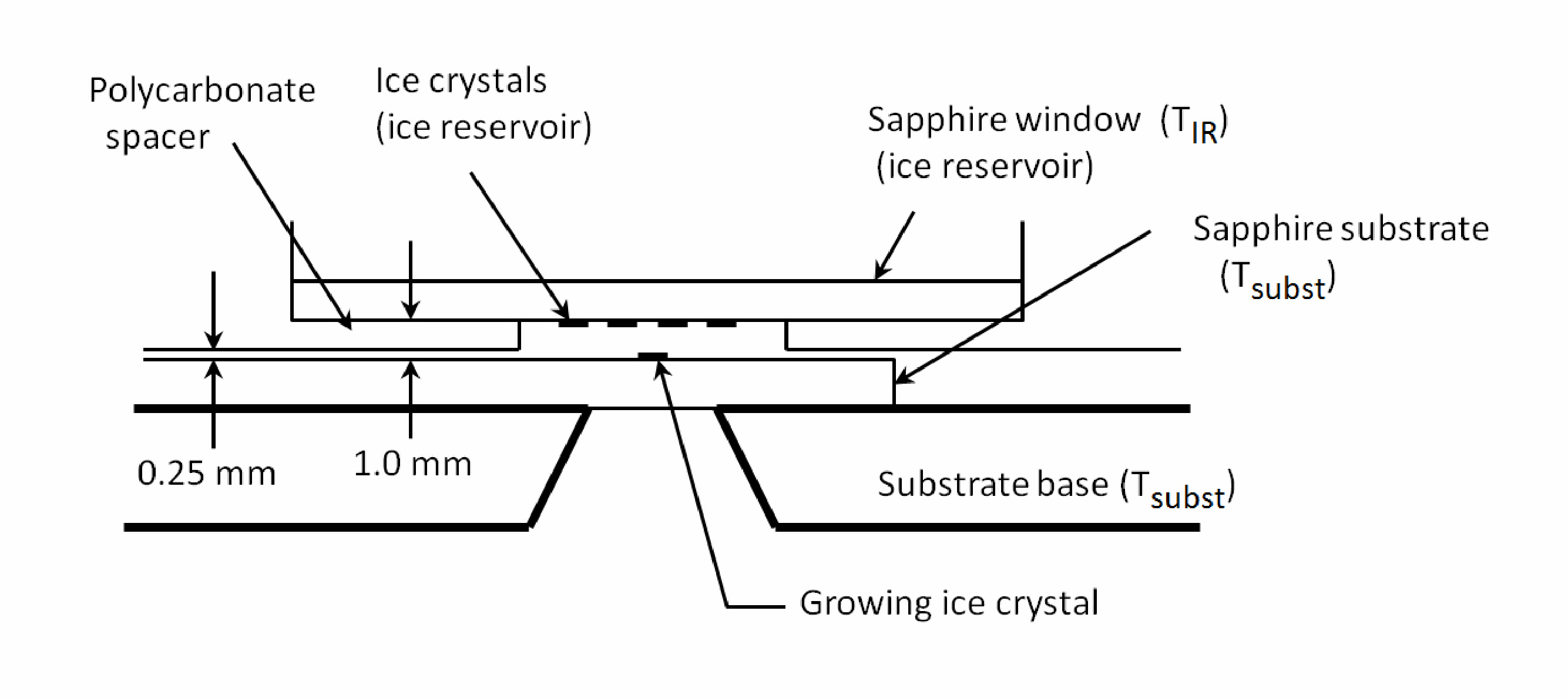}
  \caption{The experimental sub-chamber
within the inner ice growth chamber (see Figure \protect\ref{innerchamber}),
as described in the text. This sub-chamber provides an environment with
carefully controlled temperature, supersaturation, and background gas
pressure, within which we can measure ice crystal growth rates.}
  \label{subchamber}
\end{figure}

Ice growth measurements are performed in the experimental sub-chamber, shown
schematically in Figure \ref{subchamber}, that is part of the inner growth
chamber shown in Figure \ref{innerchamber}. The top of the sub-chamber is a
sapphire window with many ice crystals deposited on its lower surface, and
these ice crystals serve as an ice reservoir in our experiments. The sample
crystal rests on the sapphire substrate 1.0 mm below the ice reservoir, as
shown in Figure \ref{subchamber}. A polycarbonate spacer with an
8.1-mm-diameter hole serves as the sub-chamber walls. The sample crystal
temperature is equal to $T_{subst}$ (see Figure \ref{subchamber}), and the
supersaturation seen by the sample crystal is determined by $%
T_{IR}-T_{subst},$ as described in detail below. The 0.25-mm gap between the
substrate and the polycarbonate spacer allows the test crystal to be
positioned within the sub-chamber while diminishing the diffusive coupling
between the sub-chamber and the surrounding inner chamber.

In practice we tend to operate in one of three pressure regimes:\ 1) 740
Torr. This pressure is best for selecting crystals, changing system
parameters, etc., because the crystals react slowly; 2) 1-2 Torr. At this
pressure is it possible to easily add ice to the ice reservoir from the
moat; 3) 20 Torr. We have found that this intermediate pressure is near
optimal for measuring crystal growth rates. The pressure is high enough to
effectively isolate the experimental sub-chamber from the surrounding inner
chamber, thus reducing problems from small thermal gradients in the inner
chamber. Yet this pressure is low enough that the crystal growth is mainly
kinetics limited, thus yielding physically interesting measurements.

\subsection{The Optical Layout}

\begin{figure}[htbp] % float placement: (h)ere, page (t)op, page (b)ottom, other (p)age
  \centering
  % file name: C:/1KGLaaa/aatempfold/VIGproject/Paper1-Apparatus/arxiv/optics.gif
  \includegraphics[bb=0 0 1489 877,width=5.5in,keepaspectratio]{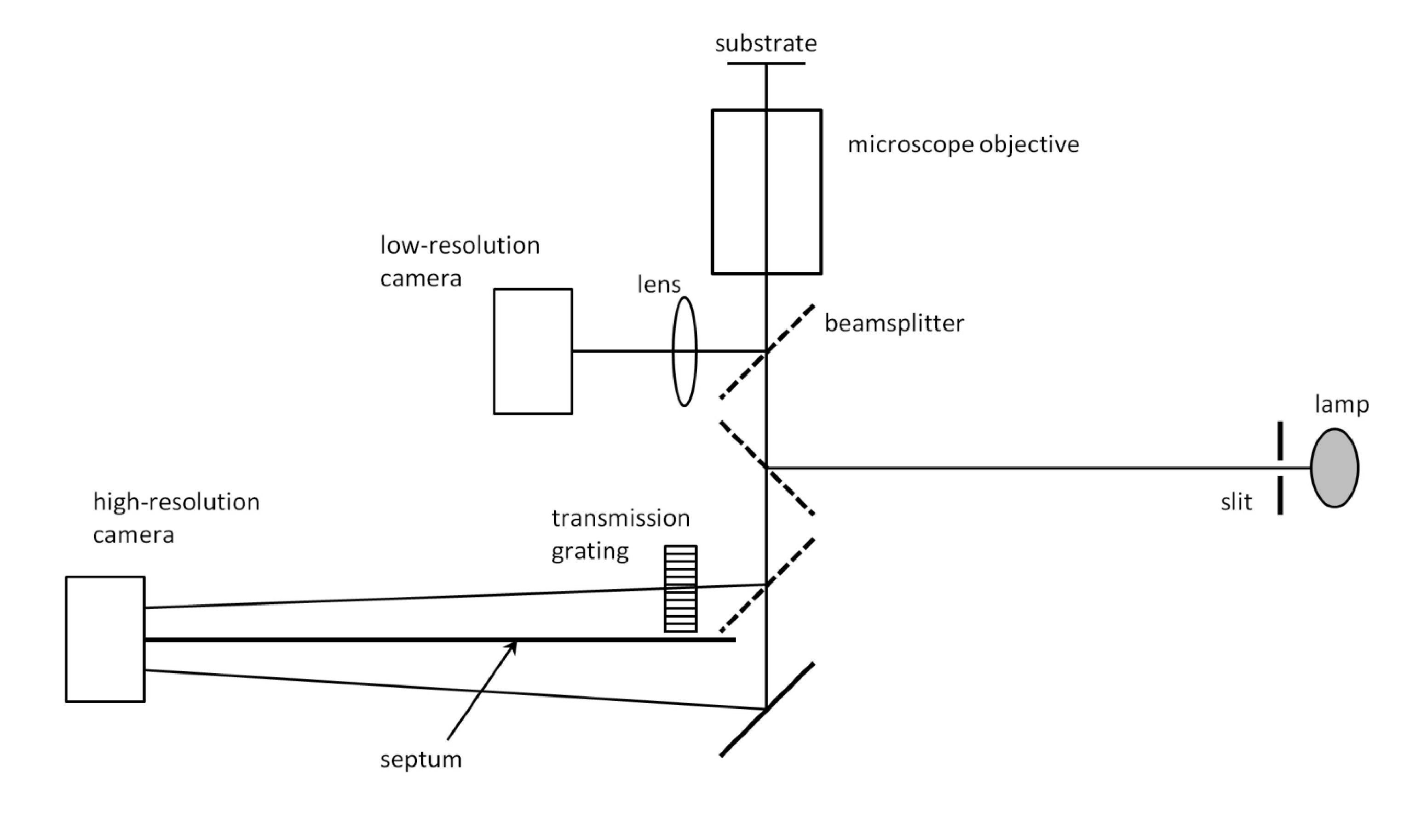}
  \caption{Optical layout (not to scale)
for imaging the growing ice crystal on the substrate and measuring its
thickness using broad-band interferometry, as described in the text.}
  \label{optics}
\end{figure}

\begin{figure}[htb] % float placement: (h)ere, page (t)op, page (b)ottom, other (p)age
  \centering
  % file name: C:/1KGLaaa/aatempfold/VIGproject/Paper1-Apparatus/arxiv/StillM134ab.gif
  \includegraphics[bb=0 0 890 668,width=4.5in,keepaspectratio]{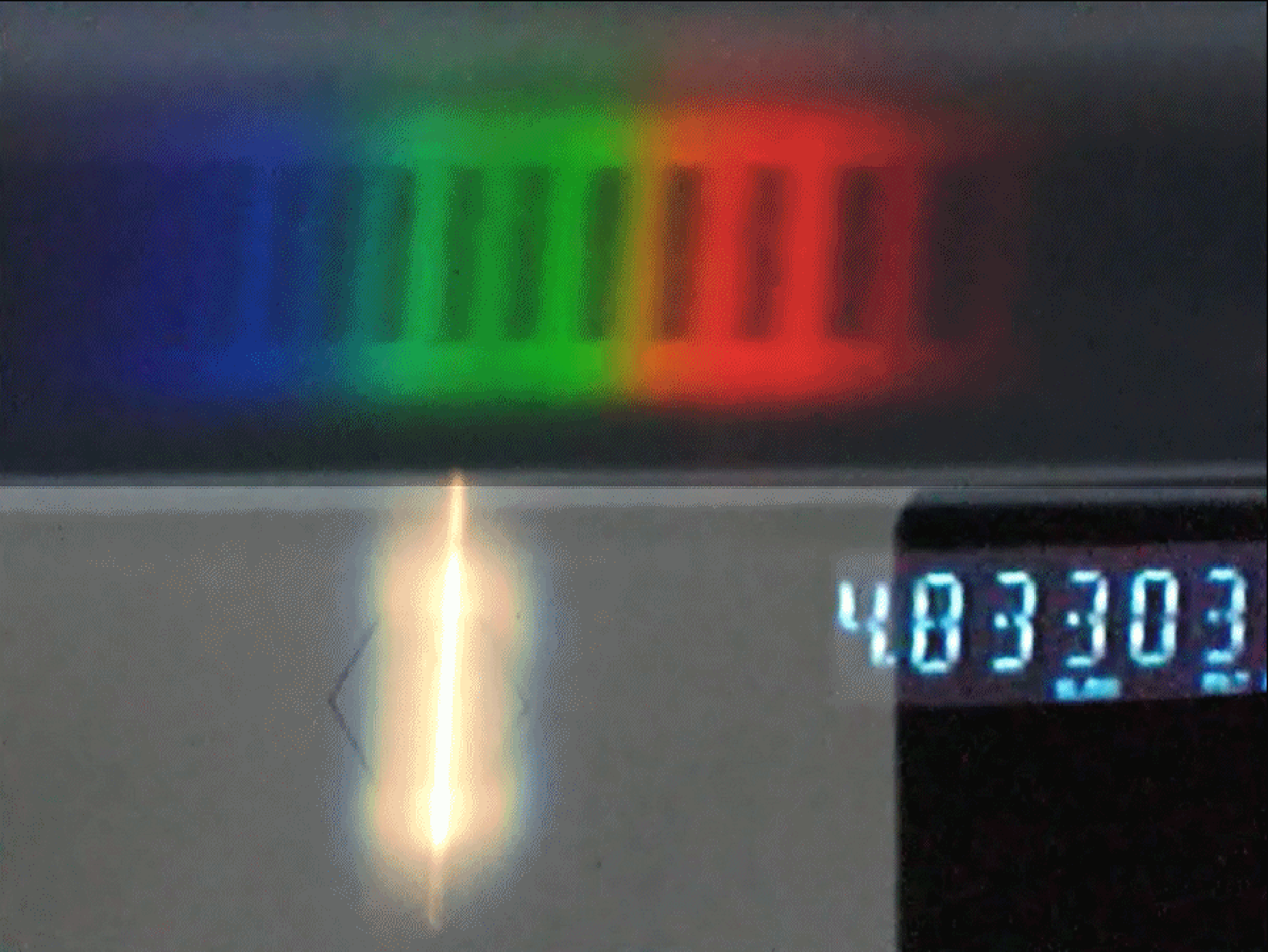}
  \caption{Sample display from the high-resolution imaging sensor (see Figure 
\protect\ref{optics}), as described in the text.}
  \label{display}
\end{figure}

The optics are set up to allow
simultaneous measurement of 1) the lateral size of the test crystal size via
direct imaging, 2) the thickness of the test crystal via broad-band
interferometry, and 3) the supersaturation via measurement of $T_{IR},$
recording all three measurements in a single video file. Figure \ref{optics}
shows a schematic diagram of the optics layout.

The microscope objective (10X Mitutoyo Plan Apo) was chosen for its 0.28
numerical aperture, giving a 1.0 $\mu $m resolving power, together with its
33.5-mm working distance, which allows ample thermal separation between the
room-temperature objective and the substrate, as shown in Figure \ref%
{innerchamber}. The crystals are illuminated from above using an external
white-light LED that shines through the ice reservoir. The upper
beamsplitter is a 10:90 (reflection:transmission) pellicle beamsplitter, and
the reflected beam gives a low-magnification image of the test region. This
image was mainly used for selecting test crystals and observing any
neighboring crystals that grew near the test crystal.

The second beamsplitter is a 50:50 pellicle beamsplitter used to input light
for the broad-band interferometer. For the source we chose a halogen bulb
with a rather broad coiled filament, using an adjustable mechanical slit to
reduce the width of the input source. An image of the bright slit is focused
by the objective onto the ice crystal, where some of the light is reflected
from the sapphire/ice interface and some is reflected by the ice/air
interface. Heating from this light is negligible. The two reflections go
back through the objective and two beamsplitters, are reflected by a third
beamsplitter (50:50), and pass through a transmission grating (70 lines/mm)
to produce a spectrum with interference fringes on the high-magnification
imaging sensor. The spacing of the fringes gives the absolute crystal
thickness, and small changes in the phase of the fringes allow accurate
measurements of growth velocities. Both cameras are Sony Alpha NEX-3
cameras, chosen for their large imaging sensors and good light sensitivity,
together with their live HDMI video outputs and HD movie recording modes.

Another portion of the light passes through all three beamsplitters and is
steered by mirror M1 onto the high-magnification sensor. A blackened septum
prevents overlap of the two incident beams, so the interferometer spectrum
is recorded by the top half of the sensor while the direct image appears on
the bottom half. Not shown in Figure \ref{optics} is a small lens and prism
in front of the high-magnification sensor that projects an image of a
television screen onto one corner of the sensor. The television is fed from
a camera viewing a precision voltmeter that measures a signal (described
below) from which $T_{IR}$ can be derived.

A sample of the final high-magnification image is shown in Figure \ref%
{display}. The bottom part of the image shows a plate-like ice crystal,
hexagonal in shape, here party obscured by the incident broad-band light
from the slit. When acquiring data, a shutter periodically blocks the light
from the interferometer slit to give a clearer direct image of the crystal.
The top part of the image shows the interference spectrum, while the digital
voltmeter output appears on the right side of the image. From a video
recording of this image we obtain the crystal size and thickness, along with
the supersaturation in the experimental sub-chamber, as a function of time.
Other experimental parameters are added via the audio channel to the same
video file.

\section{Calibration}

\subsection{Camera Image Scales}

An engraved reticle with markings spaced at 0.05 mm was placed directly on
the substrate to calibrate the image scale of both cameras. The high-res
camera was calibrated in movie mode, since data are acquired in this mode.
Note that with the Sony NEX-3 camera the absolute image size in camera mode
is different from that in movie mode; different portions of the sensor are
used in the two modes. Different movie modes (with different resolutions and
aspect ratios) also give different calibrations. Our data were all acquired
in VGA movie mode, since the image resolution was more limited by the optics
than by the sensor. During a replay of the calibration movie, PaintShopPro
was used to do a screen capture and analyze the image. The horizontal
distance between the inside edges of the movie frame was found to be 
\begin{equation}
L_{horiz}=400\pm 1\textrm{ }\mu \textrm{m}
\end{equation}

The low-res camera was calibrated in a similar fashion. The image circle,
defined by the hole in the aluminum substrate support plate directly below
the substrate, was found to have a diameter of%
\begin{equation}
D=3.1\pm 0.1\textrm{ mm}
\end{equation}

\subsection{Interferometer}

The interferometer (IFO) was calibrated by shining red and green lasers
through the lamp bulb while the camera was recording in movie mode. When the
laser alignment was good, a bright spot appeared at the location of the slit
and in the dispersed image. Both lasers were scanned in position to produce
a series of spots outlining the slit position. The spots were analyzed using
screen captures of the movie replay, and a composite image
(IFOcalibration3.jpg) was made. This image shows the various laser spots,
lines through the spots, and the superimposed IFO spectrum. A He-Ne ($%
\lambda =633$ nm) laser and a green laser pointer ($\lambda =532$ nm, a
frequency-doubled YAG laser) were used.

For data taking, we measure the fringe spacing in the orange part of the
spectrum, at $\lambda _{orange}=580$ nm as determined from the calibration
image. There is a fairly well-defined transition between red and green at
this point (see Figure \ref{display}), and the fringes are usually fairly
clear at this wavelength.

From \cite{morrison}, the reflected intensity can be written%
\[
I\sim \left[ 1+\cos \left( \frac{4\pi hn}{\lambda }\right) \right] 
\]%
where $h$ is the crystal thickness, $n$ is the index of refraction of ice,
and $\lambda $ is the light wavelength. A phase change of $2\pi $
corresponds to one IFO cycle (fringe), so%
\begin{eqnarray*}
\Delta \left( \frac{4\pi hn}{\lambda }\right) &=&2\pi \\
\left( \frac{2hn}{\lambda ^{2}}\right) \Delta \lambda &=&1 \\
h &=&\frac{\lambda ^{2}}{2n\Delta \lambda }
\end{eqnarray*}%
where $\Delta \lambda $ is the wavelength difference corresponding to a
single fringe.

To complete the calibration we use the measured line positions from
IFOcalibration3.jpg. The grating equation tells us that the dispersion of
the transmission grating is proportional to $\Delta \lambda $ (in the
small-angle approximation). The distance between the laser lines (101 nm) is
measured to be 0.2955 of the full screen width, so the full screen width
corresponds to $\Delta \lambda _{fw}=342$ nm.

With $\lambda =\lambda _{orange}=580$ nm and $n=1.31,$ this becomes%
\begin{eqnarray}
h &=&\frac{\lambda ^{2}}{2n\Delta \lambda _{fw}}\left( \frac{\Delta \lambda
_{fw}}{\Delta \lambda }\right)  \label{thickness1} \\
&=&0.375\left( \frac{L_{fw}}{\Delta L_{fringe@orange}}\right) \textrm{ }\mu 
\textrm{m}  \nonumber
\end{eqnarray}%
where $L_{fw}$ is the full-width of the movie screen and $\Delta L_{fringe}$
is the spacing between adjacent fringes at 580 nm. The largest uncertainty
in measuring the thickness of a crystal is from estimating $\Delta
L_{fringe} $ near the orange part of the spectrum, since the fringe spacing
varies with wavelength. Achieving an uncertainty of 10 percent or better is
straightforward.

The motion of the fringes is used to measure the growth velocity. From the
above we see that a single fringe passing by a fixed wavelength $\lambda $
corresponds to a thickness increase of 
\begin{eqnarray*}
\Delta \left( \frac{4\pi hn}{\lambda }\right) &=&2\pi \\
\Delta h &=&\frac{\lambda }{2n}
\end{eqnarray*}%
With a He-Ne laser incident on a crystal this gives $\Delta h_{HeNe}=242$
nm. Measuring at $\lambda =\lambda _{orange}=580$ nm gives 
\begin{equation}
\Delta h_{orange}=221\textrm{ nm}  \label{deltathick}
\end{equation}

\subsection{Supersaturation}

In the absence of a test crystal, the ice reservoir creates a
supersaturation $\sigma $ immediately above the substrate that is given by%
\begin{eqnarray*}
\sigma &=&\frac{\Delta c}{c} \\
&\approx &\frac{1}{c}\frac{dc}{dT}\Delta T \\
&\approx &\eta \Delta T
\end{eqnarray*}%
where $\Delta T=T_{IR}-T_{subst},$ $c=c_{sat}(T_{subst})$ is the saturated
water vapor pressure of ice, and $\Delta
c=c_{sat}(T_{IR})-c_{sat}(T_{subst}).$ This is a useful way to write $\sigma 
$ because $\eta $ varies only slowly with temperature. Note that a
supersaturation of 1 percent at -15 C corresponds to a temperature
difference of $\Delta T=0.11$ C.

The temperature difference $\Delta T$ is obtained from the temperature
controller output voltage for the ice reservoir $V_{IR}(T_{IR}).$ At the
beginning of each growth run we determine the $\sigma =0$ point by observing
when each crystal begins evaporating (see the evaporation tests described
below). Once the $\sigma =0$ voltage $V_{0}$ is known, the temperature
difference $\Delta T=T_{IR}-T_{subst}$ is derived from 
\begin{eqnarray*}
\Delta V_{IR} &=&\left( \frac{dV_{IR}}{dT}\right) \Delta T \\
\Delta T &=&\left( \frac{dV_{IR}}{dT}\right) ^{-1}\Delta V
\end{eqnarray*}%
where $\Delta V=\left( V_{IR}-V_{0}\right) ,$ and values for $\left(
dV_{IR}/dT\right) $ are determined empirically from a polynomial fit to
measurements of $V_{IR}(T_{IR}).$ We then have the supersaturation%
\begin{eqnarray*}
\sigma &=&\eta \left( \frac{dV_{IR}}{dT}\right) ^{-1}\Delta V \\
&=&A(T_{subst})\Delta V
\end{eqnarray*}%
Table \ref{calibration} shows values for $A(T_{subst})$ (accurate to a few
percent) as a function of temperature.

%TCIMACRO{\TeXButton{B}{\begin{table}[tbp] \centering}}%
%BeginExpansion
\begin{table}[tbp] \centering%
%EndExpansion
\begin{tabular}{|l|l|l|l|l|l|l|}
\hline
$T(C)$ & $c_{sat}/c_{solid}$ & $v_{kin}(\mu $m/s) & $\eta $ & $V_{IR}($volts)
& $dV_{IR}/dT$(mV/C) & $A$(V$^{-1})$ \\ \hline
-40 & $0.13\times 10^{-6}$ & 17 & 0.109 &  &  &  \\ 
-30 & $0.37\times 10^{-6}$ & 49 & 0.100 & 1.43 & -23.8 & 4.20 \\ 
-25 & $0.60\times 10^{-6}$ & 81 & 0.096 & 1.3118 & -23.77 & 4.04 \\ 
-20 & $0.96\times 10^{-6}$ & 131 & 0.092 & 1.1886 & -23.56 & 3.90 \\ 
-15 & $1.51\times 10^{-6}$ & 208 & 0.088 & 1.0645 & -23.14 & 3.80 \\ 
-10 & $2.33\times 10^{-6}$ & 324 & 0.085 & 0.9425 & -22.50 & 3.78 \\ 
-8 & $2.76\times 10^{-6}$ & 385 & 0.084 & 0.8950 & -22.19 & 3.79 \\ 
-5 & $3.54\times 10^{-6}$ & 496 & 0.082 & 0.8258 & -21.66 & 3.79 \\ 
-2 & $4.51\times 10^{-6}$ & 635 & 0.080 & 0.7596 & -21.06 & 3.80 \\ 
-1 & $4.88\times 10^{-6}$ & 689 & 0.079 & 0.7383 & -20.84 & 3.79 \\ \hline
\end{tabular}%
\caption{Values of various ice properties and calibration quantities as a
function of temperature.} \label{calibration}%
%TCIMACRO{\TeXButton{E}{\end{table}}}%
%BeginExpansion
\end{table}%
%EndExpansion

\subsection{Time Delay for Supersaturation Changes}

We were mindful of thermal time delays in the ice reservoir, since the
thermistor sensing $T_{IR}$ could not be located in the relevant sapphire
window. The propagation time delay between two points along a conductor is
approximately 
\[
\tau \approx \frac{c_{p}L^{2}\rho }{\kappa } 
\]%
where $c_{p}$ is the specific heat, $L$ is a characteristic length, $\rho $
is the density, and $\kappa $ is the thermal conductivity. For parts near
the ice reservoir, $\kappa \approx 30$ W/m-K for sapphire (much higher for
Aluminum), $L\approx 3$ mm, $c_{p}\approx 1000$ J/kg-K, and $\rho \approx
2000$ kg/m$^{3},$ giving $\tau \approx 0.6$ seconds. We examined the time
delay by producing sudden jumps in $T_{IR}$ and observing the subsequent
crystal growth behavior. In all cases the growth responded quickly,
paralleling the measured $T_{IR},$ as expected.

\section{Initial Measurements}

\subsection{Evaporation Measurements}

\begin{figure}[htbp] % float placement: (h)ere, page (t)op, page (b)ottom, other (p)age
  \centering
  % file name: C:/1KGLaaa/aatempfold/VIGproject/Paper1-Apparatus/arxiv/evaps.gif
  \includegraphics[bb=0 0 1397 560,width=5.5in,keepaspectratio]{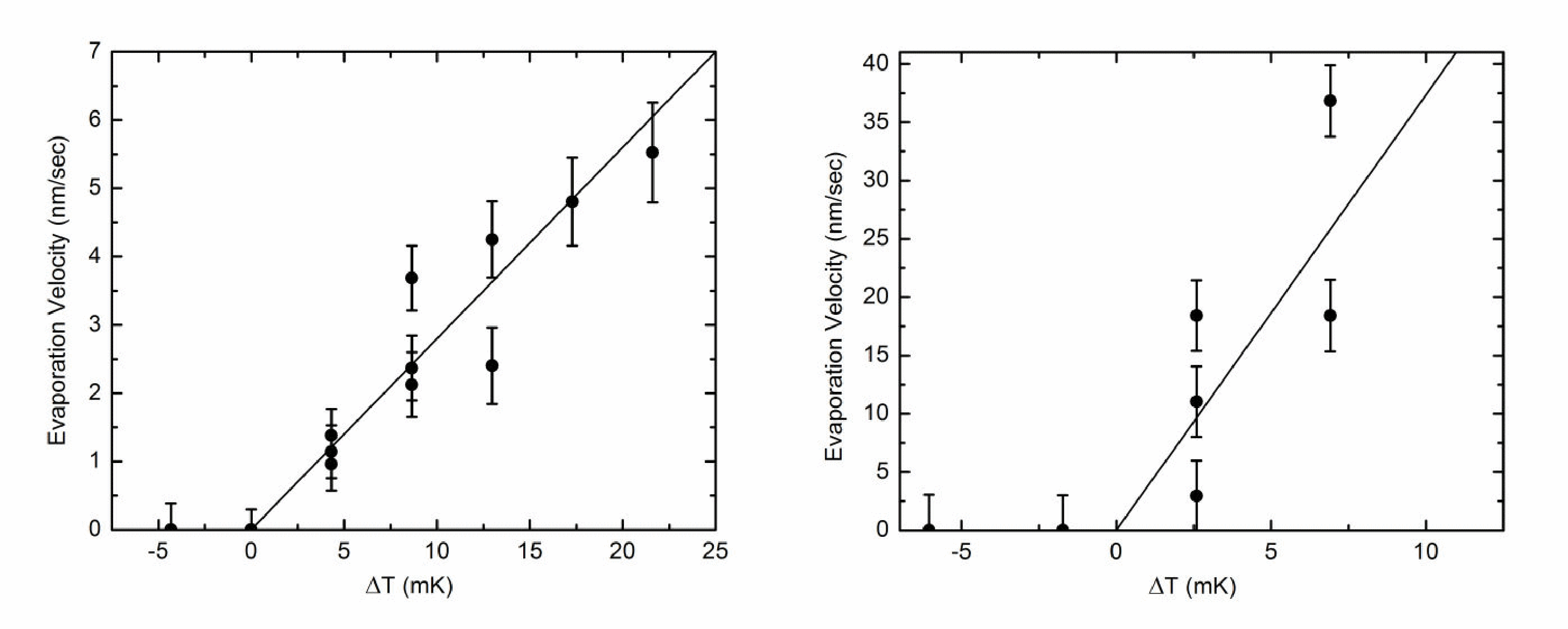}
  \caption{Measurements of evaporation
velocities for the basal facets of two ice crystals, as a function of $%
\Delta T=T_{subst}-T_{IR}$ as described in the text. The plot on the left
shows data taken at a background pressure of 740 Torr, and the plot on the
left shows data taken at 20 Torr. Lines show expected velocities for
diffusion-limited evaporation with $\protect\alpha _{diff}=0.015$ (left) and 
$\protect\alpha _{diff}=0.2$ (right).}
  \label{evaps}
\end{figure}

Once an ice crystal has been transported to the experimental chamber, we
typically first determine the saturation point (i. e. where the
supersaturation is $\sigma =0$) by slowly decreasing $\Delta T$ and
observing when the crystal begins to evaporate. This is necessary because
the individual measurements of $T_{IR}$ and $T_{subst}$ do not have
sufficient absolute accuracy to determine the $\Delta T=0$ point with the
desired precision. Data demonstrating crystal evaporation are shown in
Figure \ref{evaps}. The evaporation rates are limited by water vapor
diffusion through the surrounding air, as described by an $\alpha _{diff}$
that depends on air pressure and crystal size (see Equation \ref{diff3}). As
expected, we see in Figure \ref{evaps} that evaporation velocities are
substantially higher at lower pressures, reflecting the fact that $\alpha
_{diff}$ is inversely proportional to pressure. The first crystal in Figure %
\ref{evaps} had a diameter of 18 $\mu $m and a thickness of 3.8 $\mu $m, and
Equation \ref{diff3} gave a value of $\alpha _{diff}$ consistent with that
indicated in the data. The second crystal was roughly twice as large, and
again the calculated $\alpha _{diff}$ was consistent with that indicated in
the data. As expected, the data suggest $\alpha _{diff}<\alpha $ in all
cases, so the evaporation rates are limited essentially entirely by
diffusion.

In practice we do not acquire a great deal of evaporation data for every
crystal; instead we reduce $\Delta T$ only until a crystal shows the first
signs of evaporation, as this is typically sufficient to determine the $%
\sigma =0$ point. Shining the light from the interferometer slit on one edge
of a crystal and looking for reflection \textquotedblleft
glints\textquotedblright\ has proven to be an especially effective method
for observing small evaporation changes in real time. In this way we
estimate that we can determine the $\sigma =0$ point to a temperature
uncertainty of approximately $\Delta T=\pm 0.003$ C in most of our growth
measurements at 20 Torr, which corresponds to a supersaturation uncertainty
of $\Delta \sigma \approx \pm 0.03$ percent. Repeated evaporation tests show
that temperature drifts in the experimental chamber are typically $\Delta
T\approx \pm 0.003$ C over periods of \symbol{126}30 minutes once the
chamber has stabilized.

\subsection{Droplet Equilibrium Measurements}

We were able to test our supersaturation calibration by first increasing $%
T_{IR}$ until droplets formed and grew on the substrate, and then adjusting $%
T_{IR}$ until the droplets were neither growing nor evaporating. This
produced a known supersaturation with $\sigma =\sigma _{water}$ that we
could compare with the value calculated from our calibration. At -15 C, for
example, we measured $\sigma _{water}=15.3\pm 1.6\%,$ in agreement with the
known value of $\sigma _{water}=15.75\%$. The largest uncertainly came from
determining the stability temperature of the droplets.

\subsection{Basic Growth Measurements}

\begin{figure}[htb] % float placement: (h)ere, page (t)op, page (b)ottom, other (p)age
  \centering
  % file name: C:/1KGLaaa/aatempfold/VIGproject/Paper1-Apparatus/arxiv/BasalGrowth.gif
  \includegraphics[bb=0 0 1144 877,width=4.3in,keepaspectratio]{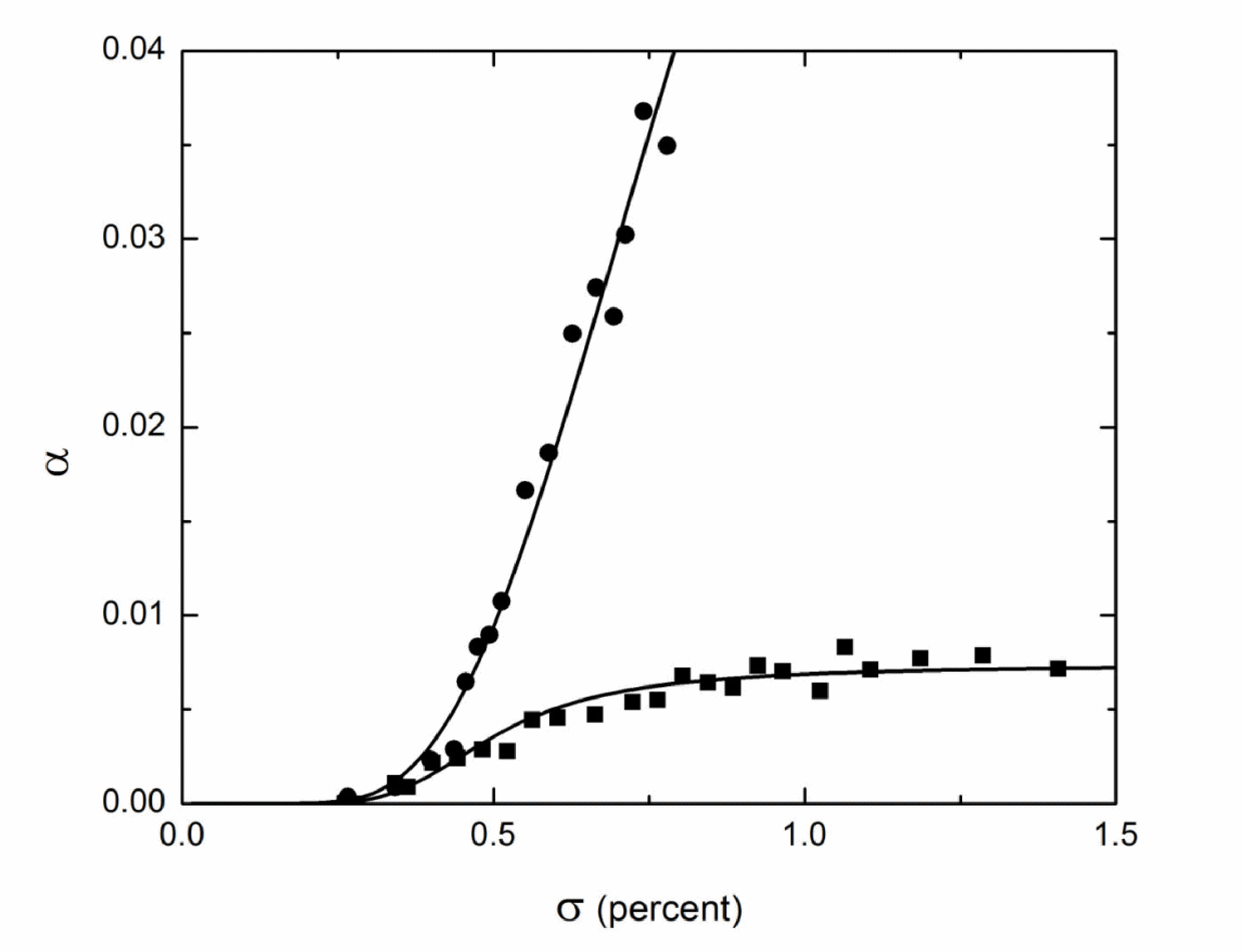}
  \caption{Measurements of the growth of
the basal facets of two ice crystals at -15 C, shown as the condensation
coefficient $\protect\alpha $ as a function of supersaturation $\protect%
\sigma .$ One crystal (dots) was grown in a background pressure of air at 20
Torr and the other (squares) was grown in a background pressure of 740 Torr.
The low-pressure crystal shows mainly kinetics-limited growth, while the
growth at high pressure is mainly limited by diffusion when the
supersaturation is high.}
  \label{basalgrowth}
\end{figure}

Figure \ref{basalgrowth} shows a typical measurement of the growth of two
ice crystals, both initially produced at -12.5 C in the outer chamber before
being transferred into the inner chamber at -15 C. In the experimental
sub-chamber, one crystal was grown in air at a background pressure of 20
Torr, while the other was grown at a pressure of 740 Torr. In both cases the
supersaturation was slowly increased by increasing $T_{IR}$ while monitoring
the interferometer output to determine the crystal thickness. From this the
growth velocity of the basal facet was derived, which in turn was used to
extract the condensation coefficient $\alpha $ as a function of the
supersaturation $\sigma .$

The basal facets of ice are well described by a nucleation-limited growth
model \cite{libbrechtreview}, and this is seen in Figure \ref{basalgrowth}
as well. For both crystals we fit the data with curves of the form%
\[
\alpha \left( \sigma \right) =\frac{A\exp \left( -\sigma _{0}/\sigma \right)
\alpha _{fit}}{A\exp \left( -\sigma _{0}/\sigma \right) +\alpha _{fit}} 
\]%
where $\alpha =A\exp \left( -\sigma _{0}/\sigma \right) $ is the intrinsic
condensation coefficient of the ice surface and $\alpha _{fit}$ is a single
fit parameter that accounts for particle diffusion to some approximation
using Equation \ref{diff2}. The fits used in Figure \ref{basalgrowth} are $%
(A,\sigma _{0},\alpha _{fit})$ = $(1,2.3,0.15)$ and $(1,2.5,0.0075)$ for the
low-pressure and high-pressure crystals, respectively. Note that $\alpha
_{fit}$ is much lower for the high-pressure crystal, reflecting the fact
that the growth is more limited by diffusion in that case. For the
low-pressure case, the diffusion term produces only a small improvement in
the fit to the data. The data are somewhat insensitive to the choice of $A,$
so this was fixed at $A=1,$ giving the physically reasonable result that the
intrinsic $\alpha \rightarrow 1$ at very high $\sigma .$

A significant result from these data, and from numerous similar crystal
measurements we have made at -15 C, is that the measured condensation
coefficients at high and low pressures are not measurably different other
than from the substantial effects of particle diffusion. This contradicts
the conclusion made in \cite{chemical}, and we now believe that the reason
for this inconsistency is an incorrect interpretation of the data in \cite%
{convection}, and we plan to reanalyze these data in a future publication.

\subsection{Measurements of $\protect\alpha _{diff}$}

\begin{figure}[ht] % float placement: (h)ere, page (t)op, page (b)ottom, other (p)age
  \centering
  % file name: C:/1KGLaaa/aatempfold/VIGproject/Paper1-Apparatus/arxiv/AlphaRadius2.gif
  \includegraphics[bb=0 0 979 751,width=3.2in,keepaspectratio]{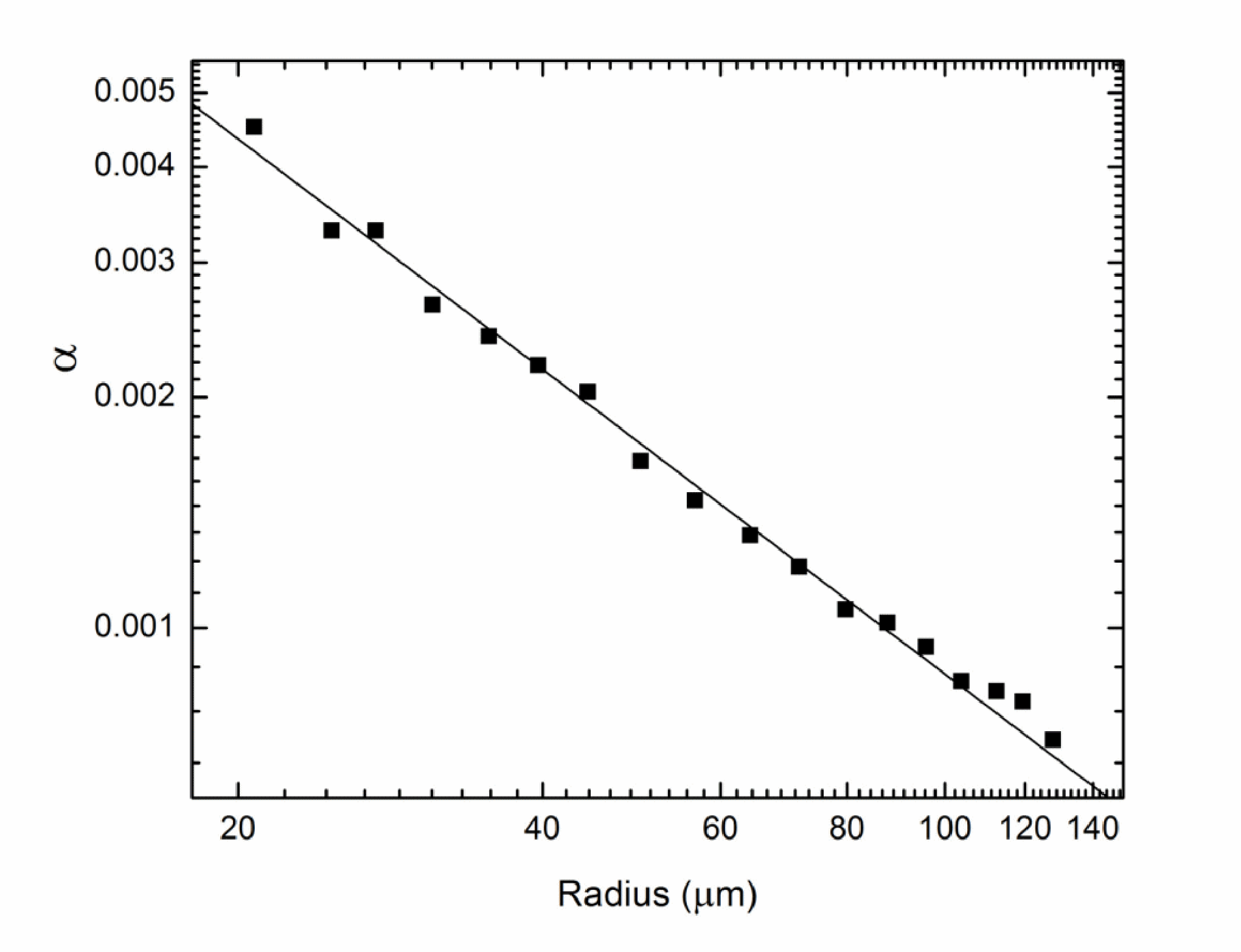}
  \caption{Measurements of the growth rate
of the top basal surface of a thin plate crystal at 740 Torr. The data
points show $\protect\alpha _{eff}=v(v_{kin}\protect\sigma _{\infty })^{-1}$
as a function of plate radius, where $v$ is the measured basal velocity. The
fit line is $\protect\alpha _{eff}=R_{0}/R$ with $R_{0}=0.087$ $\protect\mu $%
m. The data are in good agreement with a simple model for purely
diffusion-limited growth, as described in the text.}
  \label{alphadiff}
\end{figure}

It is instructive to demonstrate the $R^{-1}$ dependence of $\alpha _{diff}$
in a growing crystal, simply as a verification of our understanding of
diffusion-limited growth. For this we examined the growth of a large
plate-like crystal as it formed above the substrate at 740 Torr at -15 C,
atop an ice \textquotedblleft pedestal\textquotedblright , essentially one
half of a capped column crystal. Figure \ref{alphadiff} shows the growth
velocity expressed as $\alpha _{eff}=v(v_{kin}\sigma _{\infty })^{-1},$
where $v$ is the measured basal velocity, plotted as a function of the
measured plate radius (an average over the hexagonal structure of the
plate). For this data $\sigma _{\infty }$ was held fixed at 0.12 as the
crystal grew. The fit line is $\alpha _{eff}=R_{0}/R$ with $R_{0}=0.087$ $%
\mu $m, and we see that the data show the expected $R^{-1}$ dependence. This
simple model does not take into account the thin edges of the growing plate,
nor the geometrical differences between a flat plate and a hemisphere, and
these shortcomings likely explain the difference between $R_{0}$ and the
value found in Equation \ref{diff3}.

A remaining question in this analysis is whether the growth of this crystal
is actually diffusion-limited at all points shown in Figure \ref{alphadiff}.
From Equation \ref{diff2} we see that $\alpha (\sigma _{surf})\sigma
_{surf}=\alpha _{eff}\sigma _{\infty },$ where $\alpha _{eff}=\alpha \alpha
_{diff}/(\alpha +\alpha _{diff})$ is plotted in Figure \ref{alphadiff}.
Using $\alpha (\sigma _{surf})\approx \exp (-0.02/\sigma _{surf})$ (obtained
from measurements taken at lower pressures) we estimate that $\alpha
_{diff}\approx \alpha /20$ for the measurements in Figure \ref{alphadiff},
and thus $\alpha _{eff}\approx \alpha _{diff}$ and the growth is mainly
diffusion limited. Our overall conclusion from this exercise is that the
basal growth is well described by diffusion-limited growth when $\alpha
_{diff}<\alpha $, in reasonable agreement with expectations.

\section{Conclusions}

In summary, we have constructed an apparatus designed to make precise
measurements of ice crystal growth rates from water vapor over a range of
environmental conditions. Our particular focus was to produce exceptionally
stable and well-defined supersaturations at different temperatures and
pressures, in order to measure the intrinsic attachment coefficients for ice
growth. Detailed results from experimental measurements with this apparatus
will be reported elsewhere.

\end{document}